\documentclass[superscriptaddress, amsmath, amssymb, notitlepage, 10pt, twocolumn, notitlepage, longbibliography, aps, superscriptaddress]{revtex4-1}
\usepackage{ mathrsfs } 
\usepackage{color}
\usepackage{graphicx}
\usepackage{bmpsize}
\usepackage{dcolumn}
\usepackage[T1]{fontenc}
\usepackage{bm}
\usepackage{bbm}
\usepackage{hyperref}
\usepackage{cleveref}
\usepackage{braket}
\usepackage{mathtools}

\begin{document}
\title{Revisiting electromagnetic response of superconductors in mean-field approximation}

\author{Chang-geun Oh}
\author{Haruki Watanabe}
\affiliation{Department of Applied Physics, The University of Tokyo, Tokyo 113-8656, Japan}

\begin{abstract}
In the standard mean-field treatment of superconductors, the electron-electron interactions are assumed to be written in terms of local density operators.
However, more general interactions, such as pair-hopping interactions, may exist or may be generated in a low-energy effective Hamiltonian.
In this work, we study the effect of correlated hopping interactions toward the electromagnetic response of superconductors. When only the Hamiltonian after the mean-field approximation is provided, one cannot unambiguously determine its electromagnetic response whenever such interactions are allowed. 
This work demonstrates that such interactions induce additional terms in the current operator, leading to modifications in the Meissner weight and optical conductivities that deviate from conventional expectations. These results underscore the need for caution when incorporating gauge fields into the BdG Hamiltonian.


\end{abstract}

\maketitle

\textit{Introduction.---}
One of the most remarkable features of superconductors is the Meissner effect, which is the expulsion of an applied magnetic field from the bulk of the sample~\cite{schrieffer}.
The Bardeen--Cooper--Schrieffer (BCS) theory successfully explained the mechanism based on the mean-field approximation~\cite{schrieffer}. Although this treatment apparently breaks the $U(1)$ symmetry,  the vertex correction restores the gauge invariance of the response kernel~\cite{Nambu}.

In the study of superconductors at the mean-field level, one often starts with the Bogoliubov-de Gennes (BdG) Hamiltonian without specifying the Hamiltonian before the mean-field approximation.
The BdG Hamiltonian is not invariant under $U(1)$ phase rotation and thus its coupling to the gauge field is not uniquely determined. This results in ambiguities in the electromagnetic response of superconductors described by the BdG Hamiltonian.

\begin{figure}[t]
\includegraphics[width=\columnwidth]{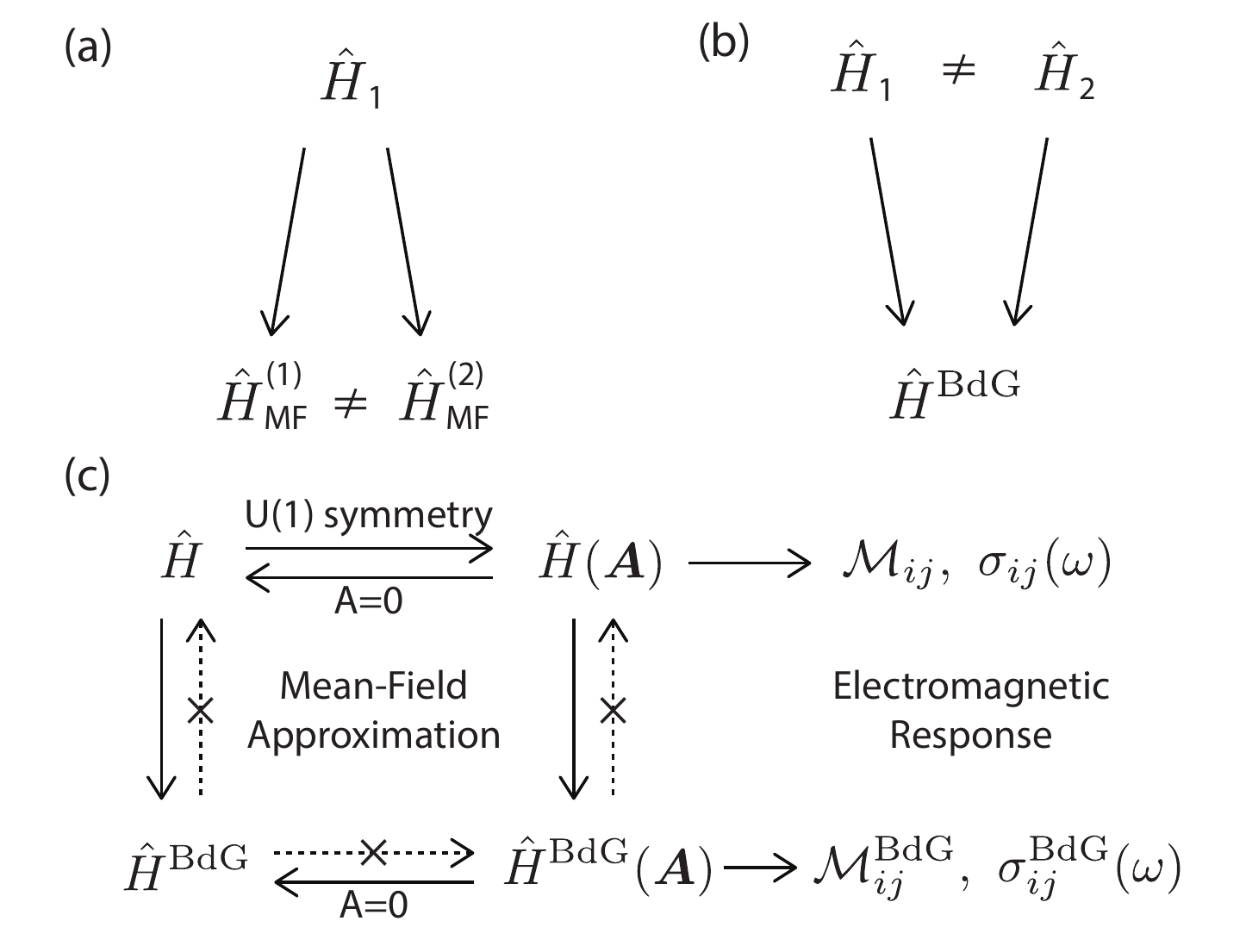}
\caption{\label{fig1} 
Schematic illustrations of (a) one-to-many relationship between microscopic model and mean-field Hamiltonian, (b) many-to-one relationship between microscopic models and BdG Hamiltonian, and (c) electromagnetic response of superconductors. Given the $U(1)$ symmetric Hamiltonian $\hat{H}$, one can unambiguously introduce the gauge field $\bm{A}$ and derive the electromagnetic response with or without the mean-field approximation. In contrast, given the BdG Hamiltonian $\hat{H}^{\mathrm{BdG}}$ alone, one cannot uniquely determine $\hat{H}^{\mathrm{BdG}}(\bm{A})$ and cannot discuss its electromagnetic response without ambiguities. 
}
\end{figure}

To see the problems in details, let us revisit the mean-field approximation from microscopic perspective.
The mean field approximation is a common technique to handle many-body interactions.
However, when employing the mean field approximation, certain ambiguities arise as shown in Fig.~\ref{fig1}.
One source of ambiguity pertains to how we define the order parameters. Depending on specific choice of order parameters, different mean-field Hamiltonians can be derived, even when applied to the same microscopic model [Fig.~\ref{fig1}(a)]. In this paper, our primary focus is on the superconducting order parameter, so this particular issue is not the central concern of our study. Another form of ambiguity arises when we only know a mean-field Hamiltonian because many microscopic models can end up with the same mean-field Hamiltonian  [Fig.~\ref{fig1}(b)]. This, in turn, leads to ambiguity in the physical observables such as electromagnetic responses. Furthermore, in the case of superconductive case, there is an additional ambiguity introduced when dealing with the incorporation of gauge fields [Fig.~\ref{fig1}(c)]. 
Let us consider superconductors described by the BdG Hamiltonian $\hat{H}^{\mathrm{BdG}}
\coloneqq
\alpha\sum_{\bm{k}}
\hat{\Psi}_{\bm{k}}^\dagger
H_{\bm{k}}^{\mathrm{BdG}}
\hat{\Psi}_{\bm{k}}+C$ with
\begin{align}
&H_{\bm{k}}^{\mathrm{BdG}}\coloneqq
\begin{pmatrix}
H_{\bm{k}}&\Delta_{\bm{k}}\\
\Delta_{\bm{k}}^\dagger&-H_{-\bm{k}}^T
\end{pmatrix}.\label{BdG1}
\end{align}
Here, $H_{\bm{k}}$ describes band dispersions for the normal phase, $\Delta_{\bm{k}}$ is the gap function that satisfies $\Delta_{-\bm{k}}=-\beta\Delta_{\bm{k}}^T$, and $C$ is a constant.
The parameters $\alpha$, $\beta$ and the Nambu spinor $\hat{\Psi}_{\bm{k}}$ are given by $\alpha=1$, $\beta=-1$, $\hat{\Psi}_{\bm{k}}\coloneqq(\hat{c}_{\bm{k}\uparrow},\hat{c}_{-\bm{k}\downarrow}^\dagger)^T$  for spinful electrons, and $\alpha=1/2$, $\beta=1$, $\hat{\Psi}_{\bm{k}}\coloneqq(\hat{c}_{\bm{k}}, \hat{c}_{-\bm{k}}^\dagger)^T$ for spinless electrons. The particle-hole symmetry $P$ satisfies $P^2=\beta\mathbbm{1}$. Note that the Hamiltonian is not invariant under $U(1)$ gauge transformation($\hat{c}_{\mathbf{k}\sigma}\to e^{i\theta}\hat{c}_{\mathbf{k}\sigma}$), because the offdiagonal terms change under this transformation. 

To examine the electromagnetic response of the superconductor, it is customary to introduce the gauge field $\bm{A}$ by replacing $H_{\bm{k}}^{\mathrm{BdG}}$ with (see, for example, Refs.~\cite{AhnNagaosa,PhysRevB.105.024308,PhysRevB.106.L220504})
\begin{align}
H_{\bm{k}}^{\mathrm{BdG}}(\bm{A})\stackrel{\mathrm{(?)}}{=}
\begin{pmatrix}
H_{\bm{k}+\bm{A}}&\Delta_{\bm{k}}\\
\Delta_{\bm{k}}^\dagger&-H_{-\bm{k}+\bm{A}}^T
\end{pmatrix}\label{HkA}
\end{align}
and define the paramagnetic current operator  and the kinetic energy operator by derivatives with respect to $\bm{A}$:
\begin{align}
&\hat{J}_i^{\mathrm{BdG}}\coloneqq\partial_{A_i}\hat{H}_{\bm{k}}^{\mathrm{BdG}}(\bm{A})|_{\bm{A}=\bm{0}},\\
&\hat{K}_{ij}^{\mathrm{BdG}}\coloneqq\partial_{A_i}\partial_{A_j}\hat{H}_{\bm{k}}^{\mathrm{BdG}}(\bm{A})|_{\bm{A}=\bm{0}}.
\end{align}
However, the $\bm{A}$-dependence of $H_{\bm{k}}^{\mathrm{BdG}}(\bm{A})$ in Eq.~\eqref{HkA} is puzzling in two ways: $\bm{A}$ does not entirely appear in the form of $\bm{k}+\bm{A}$, which normally follows by the minimum coupling, and $\Delta_{\bm{k}}$ does not depend on $\bm{A}$ even though it may have a nontrivial $\bm{k}$-dependence.


The purpose of this work is to revisit these points and clarify the subtlety behind the BdG Hamiltonian. We argue that the gap function $\Delta_{\bm{k}}$ and the constant $C$ may depend on $\bm{A}$ and contribute to the paramagnetic current operator and the kinetic operator whenever the electron-electron interactions are not solely given in terms of the electron density $\hat{n}_{\bm{x}\sigma}\coloneqq\hat{c}_{\bm{x}\sigma}^\dagger\hat{c}_{\bm{x}\sigma}$. As a consequence of these contributions, we find that the Meissner weight and the optical conductivity get modified from the standard results. 

Furthermore, the BdG Hamiltonian for $\bm{A}=\bm{0}$ is not sufficient to determine the $\bm{A}$ dependence of $\Delta_{\bm{k}}$ and $C$ unless the Hamiltonian before the mean-field approximation is provided. This means that the electromagnetic response of superconductor described by a BdG Hamiltonian alone is not completely well-defined. We clarify these points through the discussion of two illuminating examples at zero temperature $T=0$. 

\textit{$s$-wave superconductor with pair-hopping.---}
As the simplest example, let us consider a BCS type superconductor for single-band spinful electrons on $d$-dimensional hypercubic lattice:
\begin{align}
\hat{H}(\bm{A})&\coloneqq -\sum_{\bm{x}}\sum_{\sigma=\uparrow,\downarrow}\sum_{i=1}^dt(e^{-iA_i}\hat{c}_{\bm{x}+\bm{e}_i\sigma}^\dagger \hat{c}_{\bm{x}\sigma}+\text{h.c.})\notag\\
&\quad-\sum_{\bm{x}}\sum_{i=1}^d\frac{J}{2}(e^{-2iA_i}\hat{c}_{\bm{x}+\bm{e}_i\uparrow}^\dagger\hat{c}_{\bm{x}+\bm{e}_i\downarrow}^\dagger \hat{c}_{\bm{x}\downarrow}\hat{c}_{\bm{x}\uparrow}+\text{h.c.})\notag\\
&\quad-\sum_{\bm{x}}\sum_{\sigma=\uparrow,\downarrow}\mu\hat{n}_{\bm{x}\sigma}-\sum_{\bm{x}}U_0\hat{n}_{\bm{x}\uparrow}\hat{n}_{\bm{x}\downarrow},\label{H1}
\end{align}
where $\hat{c}_{\bm{k}\sigma}$'s are annihilation operators of electrons satisfying $\{\hat{c}_{\bm{k}\sigma},\hat{c}_{\bm{k}'\sigma'}^\dagger\}=\delta_{\bm{k},\bm{k}'}\delta_{\sigma,\sigma'}$, $t$ ($t>0$) is the nearest-neighbor hopping, $\mu$ ($|\mu|<2td$) is the chemical potential that might be seen as the $\nu=0$ component of the gauge field $A_\nu$, $U_0$ is the onsite density-density interaction, and $J$ is nearest-neighbor pair-hopping interaction. This model for the $U_0=A_i=0$ case is called the Penson-Kolb model~\cite{PhysRevB.33.1663,PhysRevB.59.6430} and its electromagnetic response is studied in Refs.~\cite{PhysRevB.64.104511,CZART2022126403}. We assume the periodic boundary condition with the length $L_i$ in $i$-th direction. The continuum version of this model is included in Appendix A.

The Hamiltonian has the U(1) symmetry associated with the electron density $\hat{n}_{\bm{x}}\coloneqq\sum_{\sigma=\uparrow,\downarrow}\hat{n}_{\bm{x}\sigma}$, which is necessary to fix the $\bm{A}$ dependence.
The local current operator for the link between $\bm{x}$ and $\bm{x}+\bm{e}_i$ is given by
\begin{align}
\hat{j}_{\bm{x},\bm{x}+\bm{e}_i}&\coloneqq \sum_{\sigma=\uparrow,\downarrow}t\big(ie^{-iA_i}\hat{c}_{\bm{x}+\bm{e}_i\sigma}^\dagger \hat{c}_{\bm{x}\sigma}+\text{h.c.}\big)\notag\\
&\quad+J\big(ie^{-2iA_i}\hat{c}_{\bm{x}+\bm{e}_i\uparrow}^\dagger\hat{c}_{\bm{x}+\bm{e}_i\downarrow}^\dagger \hat{c}_{\bm{x}\downarrow}\hat{c}_{\bm{x}\uparrow}+\text{h.c.}\big)
\end{align}
and satisfies the continuity equation
$
i\big[\hat{n}_{\bm{x}},\hat{H}(\bm{A})\big]=\sum_{i=1}^d\big(\hat{j}_{\bm{x},\bm{x}+\bm{e}_i}-\hat{j}_{\bm{x}-\bm{e}_i,\bm{x}}\big)
$.
The second term in the current operator originates from the pair-hopping interaction.
The model also possesses the spin rotation symmetry and, when $A_i=0$, the time-reversal symmetry and the inversion symmetry.

The superconducting order can be characterized by
\begin{align}
\phi\coloneqq\langle\hat{c}_{\bm{x}\downarrow}\hat{c}_{\bm{x}\uparrow}\rangle.
\end{align}
In this work, $\phi$ is assumed to be position-independent at least when $\bm{A}=\bm{0}$. However, $\phi$ may depend on $\bm{x}$ when $A_i\neq0$. In fact, the large gauge transformation $\hat{U}=e^{-2\pi i\sum_{i=1}^dm_i\hat{n}_{\bm{x}}x_i/L_i}$ ($m_i\in\mathbb{Z}$) maps $\phi$ to $e^{-4\pi i\sum_{i=1}^dm_ix_i/L_i}\phi$ and $A_i$ to $A_i+2\pi m_i/L_i$.  This means that, even if $\phi$ for $A_i=0$ is position-independent,  $\phi$ for $A_i=2\pi m_i/L_i\propto L_i^{-1}$ must depend on $\bm{x}$ and have the winding $m_i=(2\pi)^{-1}i\int_0^{L_i} dx_i(\phi'/|\phi'|)^*\partial_i(\phi'/|\phi'|)$. 
As we are interested in the large $L_i$ limit, for the consistency of our assumption we will keep $|A_i|$ much smaller than $2\pi /L_i$ and will set $A_i=0$ at the end of the calculation.

The Hamiltonian $\hat{H}(\bm{A})$ in Eq.~\eqref{H1} can be converted to the BdG form in Eq.~\eqref{HkA} by the mean-field approximation followed by the Fourier transformation $\hat{c}_{\bm{x}\sigma}\coloneqq V^{-1/2}\sum_{\bm{k}}\hat{c}_{\bm{k}\sigma}e^{i\bm{k}\cdot\bm{x}}$.  We find that $H_{\bm{k}}^{\mathrm{BdG}}(\bm{A})$ is given by the band dispersion $\xi_{\bm{k}}\coloneqq-\sum_{i=1}^d2t\cos k_i-\mu$ and the gap function $\Delta_{\bm{k}}$ is given by $\Delta(\bm{A})\coloneqq-U(\bm{A})\phi$ with $U(\bm{A})\coloneqq U_0+J\sum_{i=1}^d\cos(2A_i)$. The constant $C(\bm{A})=\sum_{\bm{k}}\xi_{\bm{k}+\bm{A}}+VU(\bm{A})|\phi|^2$ also depends on $\bm{A}$ whenever $J\neq0$.

When $\bm{A}=\bm{0}$, the self-consistent equation for $\phi$ at $T=0$ reads as
\begin{align}
\frac{U_{\text{tot}}}{2V}\sum_{\bm{k}}\frac{1}{E_{\bm{k}}}=1,\label{gap}
\end{align}
where 
$U_{\text{tot}}\coloneqq U_0+Jd$ is the renormalized interaction strength, 
$E_{\bm{k}}\coloneqq\sqrt{\xi_{\bm{k}}^2+|\Delta|^2}$ is the excitation energy of Bogoliubov quasi-particle,
and $\Delta\coloneqq-U_{\text{tot}}\phi$ is the gap function for $\bm{A}=\bm{0}$.
The self-consistent equation contains only $U_{\text{tot}}$, not $U_0$ nor $J$ separately. Hence, if only $\hat{H}^{\mathrm{BdG}}$ is given without $\hat{H}$, one cannot judge if $U_{\text{tot}}$ comes from the density-density interaction $U_0$ or the pair-hopping interaction $J$, as shown in Fig.~\ref{fig1}(a).

\textit{$p+ip$ topological superconductor in two dimension.---}
The above discussions are not restricted to $s$-wave superconductors.  As a more nontrivial example, let us discuss a single-band model of spinless electron
\begin{align}
\hat{H}(\bm{A})&\coloneqq -\sum_{\bm{x}}\sum_{i=1,2}t(e^{-iA_i}\hat{c}_{\bm{x}+\bm{e}_i}^\dagger \hat{c}_{\bm{x}}+\text{h.c.})\notag\\
&\quad-\sum_{\bm{x}}\frac{J}{4}
\Big[(ie^{i(A_2-A_1)}\hat{n}_{\bm{x}}\hat{c}_{\bm{x}+\bm{e}_1}^\dagger \hat{c}_{\bm{x}+\bm{e}_2}+\text{h.c.})\notag\\
&\quad\quad\quad\quad+(ie^{-i(A_1+A_2)}\hat{n}_{\bm{x}}\hat{c}_{\bm{x}+\bm{e}_2}^\dagger \hat{c}_{\bm{x}-\bm{e}_1}+\text{h.c.})\notag\\
&\quad\quad\quad\quad+(ie^{i(A_1-A_2)}\hat{n}_{\bm{x}}\hat{c}_{\bm{x}-\bm{e}_1}^\dagger \hat{c}_{\bm{x}-\bm{e}_2}+\text{h.c.})\notag\\
&\quad\quad\quad\quad+(ie^{i(A_1+A_2)}\hat{n}_{\bm{x}}\hat{c}_{\bm{x}-\bm{e}_2}^\dagger \hat{c}_{\bm{x}+\bm{e}_1}+\text{h.c.})\Big]\notag\\
&\quad-\sum_{\bm{x}}\mu\hat{n}_{\bm{x}}-\sum_{\bm{x}}\sum_{i=1,2}\frac{U_0}{2}\hat{n}_{\bm{x}}\hat{n}_{\bm{x}+\bm{e}_i},
\end{align}
where $U_0$ describes the density-density interaction and $J$ is a correlated hopping term that favors the $p+ip$ order. 
The model has both $U(1)$ symmetry and the four-fold rotation symmetry. The band dispersion is still given by $\xi_{\bm{k}}$ above. The current operator and its continuity equation are summarized in Appendix B.

Suppose that $\phi_i\coloneqq\langle\hat{c}_{\bm{x}+\bm{e}_i}\hat{c}_{\bm{x}}\rangle$ is nonzero and position independent. 
The constant $C$ for this model is given by $C(\bm{A})=(1/2)\sum_{\bm{k}}\xi_{\bm{k}+\bm{A}}-V\mathcal{E}_0(\bm{A})$ with
\begin{align}
\mathcal{E}_0(\bm{A})&\coloneqq-\frac{U_0}{2}(|\phi_1|^2+|\phi_2|^2)\notag\\
&\quad-J\cos A_1\cos A_2\,i(\phi_1^* \phi_2-\phi_2^* \phi_1).
\end{align}
When $J=0$, the relative phase between $\phi_1$ and $\phi_2$ are arbitrary, while $J>0$ favors $(\phi_1,\phi_2)=(i\phi,\phi)$ with nonzero $\phi$. This form of the gap function corresponds to the topological superconductor with the half-quantized thermal Hall conductance~\cite{PhysRevB.61.10267}.
Hence, the correlated hopping term (or some equivalent interaction) is necessary for the desired nontrivial topology.  Assuming this type of order, we find Eq.~\eqref{HkA} with $\Delta_{\bm{k}}(\bm{A})=(\sin k_1-i\sin k_2)U(\bm{A})\phi$ and $U(\bm{A})= U_0+2J\cos A_1\cos A_2$.
The self-consistent equation is modified to
\begin{align}
\frac{U_{\text{tot}}}{2V}\sum_{\bm{k}}\frac{1}{E_{\bm{k}}}\frac{\sin^2k_1+\sin^2k_2}{2}=1.
\end{align}
We stress that $\bm{A}$ \emph{does not} enter these quantities through the replacement of $\bm{k}$ with $\bm{k}+\bm{A}$, even though it \emph{is} introduced by the minimal coupling in the U(1) symmetric Hamiltonian.

\textit{Electromagnetic response.---}As we have seen, when the starting Hamiltonian with U(1) symmetry contains interactions not solely written in terms of density operators, the superconducting gap $\Delta_{\bm{k}}$ and the constant $C$ in the BdG Hamiltonian generally depend on the gauge field $\bm{A}$. Such dependence gives rise to additional terms in the current operator and the kinetic energy operator.
For example, the current operator  with a finite momentum $\bm{q}$ is given by
\begin{align}
\hat{J}_{i,\bm{q}}^{\mathrm{BdG}}&=\sum_{\bm{x}}e^{-i\bm{q}\cdot(\bm{x}+\frac{1}{2}\bm{e}_i)}\hat{j}_{\bm{x},\bm{x}+\bm{e}_i}\notag\\
&=\sum_{\bm{k}}\hat{\Psi}_{\bm{k}}^\dagger\gamma_{i,\bm{k}+\bm{q},\bm{k}}\hat{\Psi}_{\bm{k}+\bm{q}}+\delta_{\bm{q},\bm{0}}\partial_{A_i}C(\bm{A})|_{\bm{A}=\bm{0}}
\end{align}
with $\gamma_{i,\bm{k}+\bm{q},\bm{k}}\coloneqq v_{i,\bm{k}+\frac{\bm{q}}{2}}\sigma_0-2J\phi\sin\tfrac{q_i}{2}i\sigma_2$, which  contains the contribution from the pair-hopping interaction in the off-diagonal part, in addition to the standard band velocity term $v_{i,\bm{k}}\coloneqq\partial_{k_i}\xi_{\bm{k}}$ in the diagonal part. On the other hand, the charge density operator
\begin{align}
\hat{J}_{0,\bm{q}}^{\mathrm{BdG}}&=\sum_{\bm{x}}e^{-i\bm{q}\cdot\bm{x}}\hat{n}_{\bm{x}}=\sum_{\bm{k}}\hat{\Psi}_{\bm{k}}^\dagger\gamma_{0,\bm{k}+\bm{q},\bm{k}}\hat{\Psi}_{\bm{k}+\bm{q}}+\delta_{\bm{q},\bm{0}}V
\end{align}
with $\gamma_{0,\bm{k}+\bm{q},\bm{k}}\coloneqq\sigma_3$ is not affected by $J$.  In the reminder of this work we discuss the physical consequences of these additional terms using the spinful electron model in Eq.~\eqref{H1}. 
Without loss of generality, we set $\Delta$ to be real using the U(1) symmetry. 

\textit{Meissner Weight.---}Let us consider the linear response kernel of the current operator toward the gauge field with a frequency $\omega=q_0$ and a momentum $\bm{q}$:
\begin{align}
j_\mu(q)=\sum_{\nu=0}^d\mathcal{K}_{\mu\nu}(q)A_\nu(q).\label{defkernel}
\end{align}
Here and hereafter we write $q=(q_0,\bm{q})$ for short. According to the linear response theory, the response kernel is given by
\begin{align}
\mathcal{K}_{\mu\nu}^{\mathrm{BdG}}(q)=\mathcal{M}^{\mathrm{BdG}}_{\mu\nu}+\mathcal{R}_{\mu\nu}^{\mathrm{BdG}}(q),\label{BdGkernel}
\end{align}
where
\begin{align}
\mathcal{R}_{\mu\nu}^{\mathrm{BdG}}(q)\coloneqq-\frac{i}{V}\int dt e^{i\omega t}\theta(t)\langle [\hat{J}_{\mu,\bm{q}}^{\mathrm{BdG}}(t),\hat{J}_{\nu,-\bm{q}}^{\mathrm{BdG}}(0)]\rangle
\end{align}
is the retarded current correlation function in the mean-field approximation and
\begin{align}
\mathcal{M}_{ij}^{\mathrm{BdG}}\coloneqq\frac{1}{V}\langle \hat{K}_{ij}^{\mathrm{BdG}}\rangle
\end{align}
is the diamagnetic contribution giving the Meissner weight. We find 
\begin{align}
\mathcal{M}_{ij}^{\mathrm{BdG}}=\frac{1}{V}\sum_{\bm{k}}\langle\hat{n}_{\bm{k}}\rangle\partial_{k_i}\partial_{k_j}\xi_{\bm{k}}-|\phi|^2\partial_{A_i}\partial_{A_j}U(\bm{A})|_{\bm{A}=\bm{0}}\label{resultMW}
\end{align}
and $\mathcal{M}_{\mu\nu}^{\mathrm{BdG}}=0$ if $\mu$ or $\nu$ is $0$.
This result also applies to the spinless model and hence generalizes the result in Refs.~\cite{PhysRevB.64.104511,CZART2022126403}. 
In addition to the usual term associated with the electron density $\langle\hat{n}_{\bm{k}}\rangle=\alpha(1-\xi_{\bm{k}}/E_{\bm{k}})$ and the band curvature $\partial_{k_i}\partial_{k_j}\xi_{\bm{k}}$, the second term arises from the $\bm{A}$ dependence of $\Delta_{\bm{k}}$ and $C$. If the standard form in Eq.~\eqref{HkA} were assumed instead, this term would be missed. This effect might be measured through the penetration depth $\lambda$ of an external magnetic field.

\textit{Response kernel with vertex correction.---}
As is well-known, the kernel $\mathcal{K}^{\mathrm{BdG}}_{\mu\nu}(q)$ in Eq.~\eqref{BdGkernel} in the mean-field treatment does not respect the gauge invariance. That is, the induced current $\sum_{\nu=0}^d\mathcal{K}^{\mathrm{BdG}}_{\mu\nu}(q)A_\nu(q)$ is not invariant under the gauge transformation $A_0(q)\rightarrow A_0(q)-\omega \chi(q)$ and $A_i(q)\rightarrow A_i(q)+2\sin\tfrac{q_i}{2}\, \chi(q)$. 
To obtain the gauge-invariant optical conductivity $\sigma_{ij}(q)$ towards the electric field $E_j(q)=-i\omega A_j(q)$, let us take into account the vertex correction following the steps in Refs.~\cite{Nambu,schrieffer}.

\begin{figure}[t]
\includegraphics[width=\columnwidth]{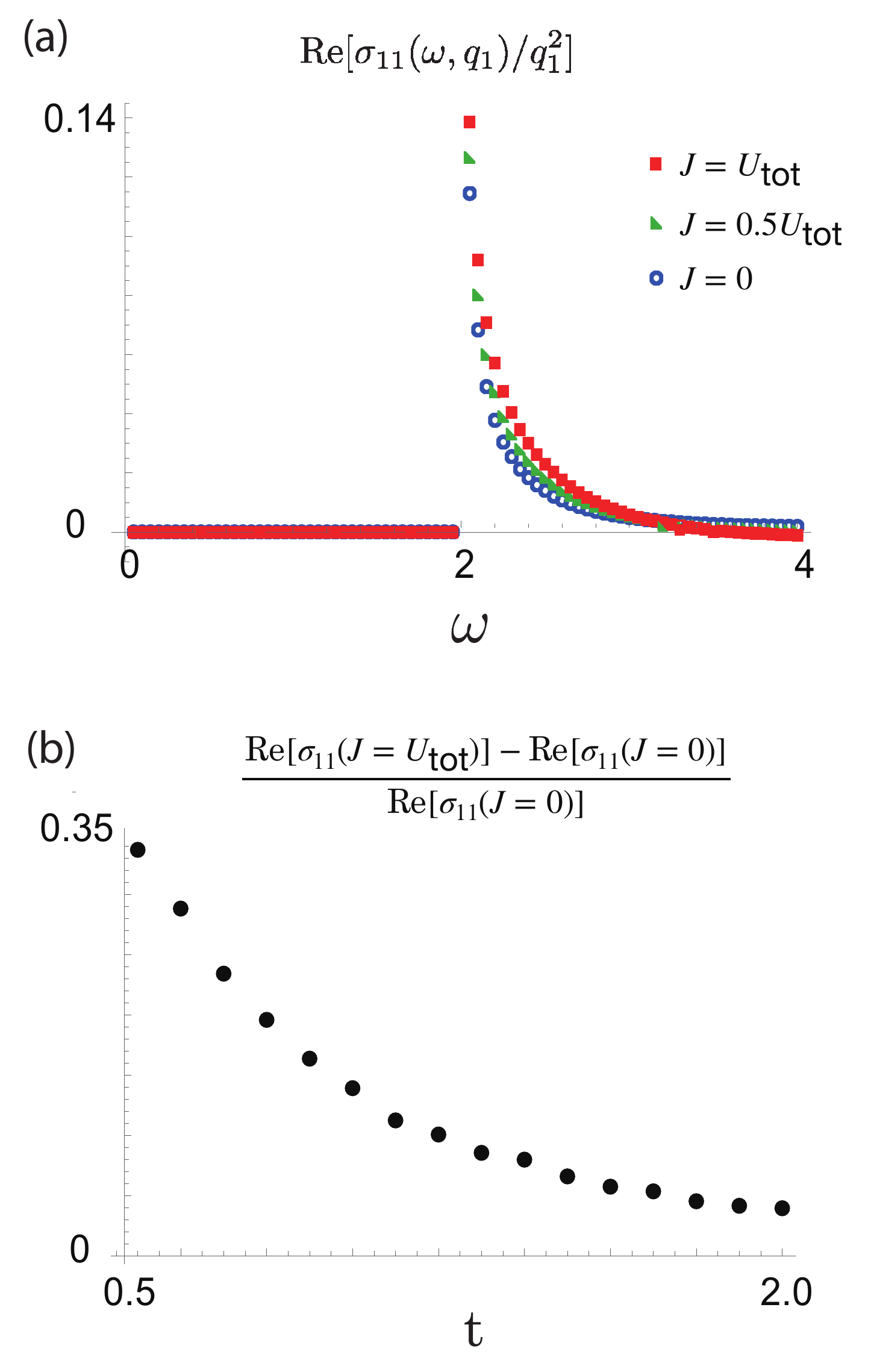}
\caption{\label{fig3}(a) The optical conductivities $\mathrm{Re}[\sigma_{11}(\omega,q_1)]$ towards nonuniform field with the vertex correction for various $J$. Here, we used $t=0.6$, $\mu=0$, $\Delta=1$. We expand $\mathrm{Re}[\sigma_{11}(\omega,q_1)]$ in the Taylor series of $q_1^2$ and here we show the coefficient of the $q_1^2$ term.  Red, green, and blue markers represent $J=U_{\text{tot}}$, $J=0.5 U_{\text{tot}}$, and $J=0$. (b) The ratio of the difference of conductivity between $J=U_\text{tot}$ and $J=0$ as a function of $t$ at $\omega = 2.1$.
}
\end{figure}

\begin{figure*}[t]
\includegraphics[width=\textwidth]{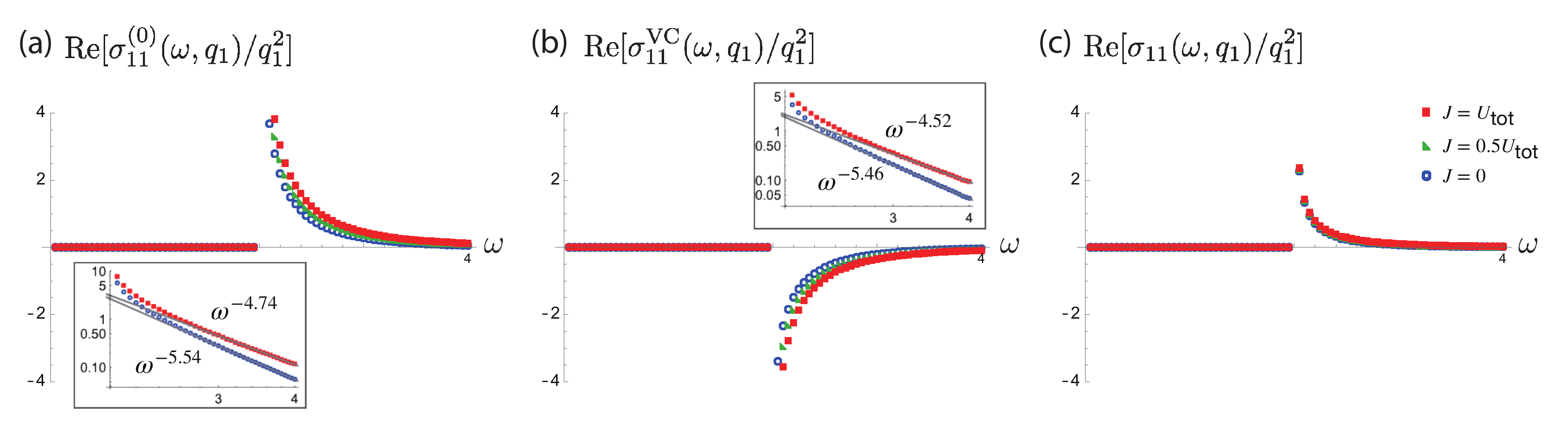}
\caption{\label{fig4}
The optical conductivities $\mathrm{Re}[\sigma_{11}(\omega,q_1)]$ towards nonuniform field with [(c)] or without [(a)] the vertex correction [(b)] for $t=1.5$, $\mu=0$, $\Delta=1$ ($U_{\text{tot}}$ is fixed by the self-consistent equation) in two dimensional system. We expand $\mathrm{Re}[\sigma_{11}(\omega,q_1)]$ in the Taylor series of $q_1^2$ and here we show the coefficient of the $q_1^2$ term.  Red squares, green triangle, and blue circles are for $J= U_{\text{tot}}$,   $J=0.5U_{\text{tot}}$, and $J=0$, respectively.
The insets in panels (a)-(b) are the log-log plots of the absolute value $|\mathrm{Re}[\sigma_{11}(\omega,q_1)/q_1^2]|$. Gray lines are obtained by fitting to determine the power of the decay $\mathrm{Re}[\sigma_{11}(\omega,q_1)/q_1^2]\propto\omega^{-n}$. 
}
\end{figure*}

First, we define the vertex function $\Gamma_\mu$ by
\begin{align}
&\langle \mathcal{T}\hat{n}_{\bm{z}}(t_z)\hat{\Psi}_{\bm{x}}(t_x)\hat{\Psi}_{\bm{y}}^\dagger(t_y)\rangle\notag\\
&=-\sum_{\bm{x}',\bm{y}'}\int dt_x'dt_y'G(x,x')\Gamma_0(x',y',\bm{z},t_z)G(y',y)
\end{align}
and
\begin{align}
&\langle \mathcal{T}\hat{j}_{\bm{z},\bm{z}+\bm{e}_i}(t_z)\hat{\Psi}_{\bm{x}}(t_x)\hat{\Psi}_{\bm{y}}^\dagger(t_y)\rangle\notag\\
&=-\sum_{\bm{x}',\bm{y}'}\int dt_x'dt_y'G(x,x')\Gamma_i(x',y',\bm{z}+\tfrac{1}{2}\bm{e}_i,t_z)G(y',y).
\end{align}
Here, 
\begin{align}
G_q\coloneqq \int dte^{i\omega t}(-i)\langle \mathcal{T}\hat{\Psi}_{\bm{q}}(t)\hat{\Psi}_{\bm{q}}^\dagger\rangle
=\frac{\omega\sigma_0+\xi_{\bm{q}}\sigma_3+\Delta\sigma_1}{\omega^2-E_{\bm{q}}^2+i\delta}
\end{align}
is the time-ordered Green function.
The continuity equation for the current operator~\cite{Nambu,schrieffer} implies the generalized Ward identity (GWI):
\begin{align}
\sum_{i=1}^d2\sin\tfrac{q_i}{2}\Gamma_{i,k+q,k}-\omega\Gamma_{0,k+q,k}=\sigma_3G_k^{-1}-G_{k+q}^{-1}\sigma_3.\label{Wardeq}
\end{align}
For reader's convenience, we review the derivation in Refs.~\cite{Nambu,schrieffer} in Appendix A. The vertex function can be obtained by solving the Bethe--Salpeter equation
\begin{align}
&\Gamma_{\mu,k+q,k}=\gamma_{\mu,\bm{k}+\bm{q},\bm{k}}^{(0)}+\mathcal{C}_{\mu,q},\\
&\mathcal{C}_{\mu,q}\coloneqq U_{\text{tot}}\int\frac{dk_0}{2\pi i}\frac{1}{V}\sum_{\bm{k}}\sigma_3G_{k+q}\Gamma_{\mu,k+q,k}G_{k}\sigma_3
\end{align}
with $\gamma_{0,\bm{k}+\bm{q},\bm{k}}^{(0)}=\sigma_3$, $\gamma_{i,\bm{k}+\bm{q},\bm{k}}^{(0)}\coloneqq v_{i,\bm{k}+\frac{\bm{q}}{2}}\sigma_0$ and $v_{i,\mathbf{k}} \coloneqq \partial_{k_i}\xi_{\mathbf{k}}$. 

Once $\Gamma_\mu$ is obtained, the time-ordered current correlation function $\mathcal{P}_{\mu\nu}(q)$ can be expressed as
\begin{align}
&\mathcal{P}_{\mu\nu}(q)\coloneqq -\frac{i}{V}\int dt e^{i\omega t}\langle \mathcal{T}\hat{J}_{\mu,\bm{q}}(t)\hat{J}_{\nu,-\bm{q}}(0)\rangle\notag\\
&\quad=\int \frac{dk_0}{2\pi i}\frac{1}{V}\sum_{\bm{k}}\text{tr}[\gamma_{\mu,\bm{k}+\bm{q},\bm{k}}
G_{k+q}\Gamma_{\nu,k+q,k}G_{k}]e^{ik_0\delta}.\label{defGamma}
\end{align}
Then the retarded correlation function $\mathcal{R}_{\mu\nu}(q)$ is given by $\mathrm{Re}\mathcal{R}_{\mu\nu}(q)=\mathrm{Re}\mathcal{P}_{\mu\nu}(q)$ and $\mathrm{Im}\mathcal{R}_{\mu\nu}(q)=\mathrm{sgn}\,\omega\,\mathrm{Im}\mathcal{P}_{\mu\nu}(q)$ (see Appendix C). The optical conductivity $\sigma_{ij}(q)$ including the vertex correction is then given by
\begin{align}
\sigma_{ij}(q)=\frac{i}{\omega}\mathcal{K}_{ij}(q)=\frac{i}{\omega}[\mathcal{M}_{ij}^{\mathrm{BdG}}+\mathcal{R}_{ij}(q)].\label{oc}
\end{align}

Using the GWI \eqref{Wardeq} and the self-consistent equation \eqref{gap}, we have
\begin{align}
&\sum_{j=1}^d \mathcal{P}_{ij}(q)(2\sin\tfrac{q_j}{2})-\mathcal{P}_{i0}(q)\omega\notag\\
&=\int \frac{dk_0}{2\pi i}\frac{1}{V}\sum_{\bm{k}}\text{tr}[(\sigma_3\gamma_{i,\bm{k}-\bm{q},\bm{k}}-\gamma_{i,\bm{k},\bm{k}+\bm{q}}\sigma_3)G_k]\notag\\
&= -\sum_{j=1}^d\mathcal{M}_{ij}^{\mathrm{BdG}}(2\sin\tfrac{q_j}{2}),
\end{align}
implying that the gauge invariance is restored: $\sum_{j=1}^d \mathcal{K}_{ij}(q)(2\sin\tfrac{q_j}{2})-\mathcal{K}_{i0}(q)\omega=0$. 

Figures~\ref{fig3} and \ref{fig4} show our numerical results for the optical conductivity $\sigma_{ij}(q)$ in Eq.~\eqref{oc} for $q_2=\cdots=q_d=0$. In single-band models with inversion and the time-reversal symmetry, the optical conductivity vanishes when $\bm{q}=0$~\cite{AhnNagaosa}. Hence, here we assume spatially modulating field with $q_1\neq0$ and focus on the coefficient of $q_1^2$ in the Taylor series of $\mathrm{Re}[\sigma_{11}(\omega,q_1)]$  with respect to $q_1$. Here, we fix $\Delta=1$, and $U_{\text{tot}}$ is fixed by the self-consistent equation.

We found that nonzero $J$ have a significant impact on the conductivity when the normal velocity $v_{i,\mathbf{k}}$ is small, indicating that the band width of the normal state or $t$ is small compared to $J$. In Fig.~\ref{fig3}(a) we present a comparison of the results for three cases: $J=0$, $J=0.5 U_{\text{tot}}$, and $J= U_{\text{tot}}$, while keeping $U_{\text{tot}}$, $t$, and $\Delta$ constant. Additionally, we examine the $t$-dependence of the ratio of the conductivity difference between $J= U_{\text{tot}}$ and $J=0$ at $\omega=2.1$ in Fig.~\ref{fig3}(b). The results show that when the normal state band is relatively flat in comparison to $\Delta$, the nonzero $J$ have a pronounced impact on conductivity. Recently, there has been a growing interest in models featuring flat or quasi-flat band superconductors~\cite{balents2020superconductivity,liang2017band,kopnin2011high}. This finding underscores the need for caution when applying standard practices to address electromagnetic response in such models.
 
Figure~\ref{fig4} shows the results for the $J=0.5 U_{\text{tot}}$ case (red squares) and the $J=0$ case (blue circles) with the large band width of the normal state compared to $J$ in two and three dimensions.
We found that nonzero $J$ significantly affect the bare conductivities [(a),(d)] and the vertex corrections  [(b),(e)] separately, although such differences are mostly suppressed in the sum $\sigma_{ij}(q)\coloneqq\sigma_{ij}^{(0)}(q)+\sigma_{ij}^{\text{VC}}(q)$ [(c),(f)]. In particular, the contribution from $J$ usually decay with smaller power of $\omega$ as shown in the insets. However, the cancellation is not perfect and there are still finite differences originating from nonzero $J$.

\textit{Conclusion.---}
In this work, we revisited the electromagnetic response of superconductors.
In general, the correspondence of the microscopic models and the BdG Hamiltonians after the mean-field approximation is ``many to one'': in the case of the spinful electron model in Eq.~\eqref{H1}, as far as the renormalized parameter $U_{\text{tot}}=U_0+Jd$ is fixed, models with different choices of on-site Coulomb interaction $U_0$ and the pair-hopping interaction $J$ lead to the same BdG Hamiltonian. However, we found that the Meissner weight and optical conductivities are sensitive to the specific value of $J$, as shown in Eq.~\eqref{resultMW}, Figs.~\ref{fig3} and \ref{fig4}. This means that the response toward U(1) gauge field has ambiguities unless the microscopic model with U(1) symmetry is provided. Our results call for caution in the standard practice in introducing the gauge-field $\bm{A}$ to the BdG Hamiltonian as in Eq.~\eqref{HkA}.

\begin{acknowledgements}
We thank Seishiro Ono, Yohei Fuji, Kazuaki Takasan, Naoto Tsuji, and Rina Tazai for useful discussions.
The work of H.W. is supported by JSPS KAKENHI Grant No.~JP20H01825 and JP21H01789.
\end{acknowledgements}

\bibliography{ref.bib}

\onecolumngrid

\clearpage

\appendix
\section{spinful model}
\label{app:spinful}

\subsection{Model}
The Hamiltonian is defined as
\begin{align}
\hat{H}(\bm{A})&\coloneqq -\sum_{\bm{x}}\sum_{\sigma=\uparrow,\downarrow}\sum_{i=1}^dt(e^{-iA_i}\hat{c}_{\bm{x}+\bm{e}_i\sigma}^\dagger \hat{c}_{\bm{x}\sigma}+\text{h.c.})-\sum_{\bm{x}}\sum_{\sigma=\uparrow,\downarrow}\mu\hat{n}_{\bm{x}\sigma}\notag\\
&\quad-\sum_{\bm{x}}U_0\hat{n}_{\bm{x}\uparrow}\hat{n}_{\bm{x}\downarrow}-\sum_{\bm{x}}\sum_{i=1}^d\frac{J}{2}(e^{-2iA_i}\hat{c}_{\bm{x}+\bm{e}_i\uparrow}^\dagger\hat{c}_{\bm{x}+\bm{e}_i\downarrow}^\dagger \hat{c}_{\bm{x}\downarrow}\hat{c}_{\bm{x}\uparrow}+\text{h.c.}).
\end{align}
In terms of Fourier components, the interaction term of the Hamiltonian can be written as
\begin{align}
&-\sum_{\bm{x}}U_0\hat{n}_{\bm{x}\uparrow}\hat{n}_{\bm{x}\downarrow}-\sum_{\bm{x}}\sum_{i=1}^d\frac{J}{2}(e^{-2iA_i}\hat{c}_{\bm{x}+\bm{e}_i\uparrow}^\dagger\hat{c}_{\bm{x}+\bm{e}_i\downarrow}^\dagger \hat{c}_{\bm{x}\downarrow}\hat{c}_{\bm{x}\uparrow}+\text{h.c.})\notag\\
&=-\frac{1}{V}\sum_{\bm{q}}\Big[U_0+J\sum_{i=1}^d\cos(q_i+2A_i)\Big]\left(\sum_{\bm{k}'}\hat{c}_{\bm{k}'+\bm{q}\uparrow}^\dagger \hat{c}_{-\bm{k}'\downarrow}^\dagger\right)\left(\sum_{\bm{k}}\hat{c}_{-\bm{k}\downarrow}\hat{c}_{\bm{k}+\bm{q}\uparrow}\right).
\end{align}

To obtain the continuum limit of the model, one can replace $\hat{c}_{\bm{x}+\bm{e}_i\sigma}$ by $\hat{c}_{\bm{x}+a\bm{e}_i\sigma}$ and $A_i$ by $aA_i$ and perform the Taylor expansion with respect to the lattice constant $a$. One finds
\begin{align}
a^d\hat{H}&\simeq \int d^dx\sum_{\sigma=\uparrow,\downarrow}\Big(ta^2(\bm{\nabla}-i\bm{A})\hat{c}_{\bm{x}\sigma}^\dagger\cdot(\bm{\nabla}+i\bm{A})\hat{c}_{\bm{x}\sigma}-(\mu+2t)\hat{n}_{\bm{x}\sigma}\Big)\notag\\
&\quad-\int d^dx(U_0+J)\hat{n}_{\bm{x}\uparrow}\hat{n}_{\bm{x}\downarrow}+\int d^dx\frac{Ja^2}{2}(\bm{\nabla}-2i\bm{A})(\hat{c}_{\bm{x}\uparrow}^\dagger\hat{c}_{\bm{x}\downarrow}^\dagger)\cdot(\bm{\nabla}+2i\bm{A})(\hat{c}_{\bm{x}\downarrow}\hat{c}_{\bm{x}\uparrow}).
\end{align}
This model is fully rotation symmetric and might be easier to deal with. However, in this work we discuss only the lattice model for the consistency.

\subsection{Fourier transformation of the current operator}
The current operator with a finite momentum $\bm{q}$ is defined by the Fourier transformation
\begin{align}
&\hat{J}_{0,\bm{q}}\coloneqq \hat{N}_{\bm{q}}=\sum_{\bm{x}}\sum_{\sigma=\uparrow,\downarrow}e^{-i\bm{q}\cdot\bm{x}}\hat{n}_{\bm{x}\sigma},\\
&\hat{J}_{i,\bm{q}}\coloneqq \sum_{\bm{x}} e^{-i\bm{q}\cdot(\bm{x}+\frac{1}{2}\bm{e}_i)}\hat{j}_{\bm{x},\bm{x}+\bm{e}_i}.
\end{align}
The continuity equation becomes
\begin{align}
[\hat{N}_{\bm{q}},\hat{H}]=\sum_{i=1}^d
2\sin\tfrac{q_i}{2}
\hat{J}_{i,\bm{q}}.
\end{align}
If one also performs a Fourier transformation with respect to $t$ as
\begin{align}
\hat{N}_{q}\coloneqq \int dt e^{i\omega t}e^{i\hat{H}t}\hat{N}_{\bm{q}}e^{-i\hat{H}t},
\end{align}
then $[\hat{N}_{q},\hat{H}]$ should be replaced with $\omega\hat{N}_{q}$. After the mean-field approximation, we find 
\begin{align}
&\hat{J}_{\mu,\bm{q}}=\sum_{\bm{p}}\hat{\Psi}_{\bm{p}}^\dagger\gamma_{\mu,\bm{p}+\bm{q},\bm{p}}\hat{\Psi}_{\bm{p}+\bm{q}},\\
&\gamma_{0,\bm{k}+\bm{q},\bm{k}}\coloneqq\sigma_3,\quad\gamma_{i,\bm{k}+\bm{q},\bm{k}}\coloneqq v_{i,\bm{k}+\frac{\bm{q}}{2}}\sigma_0-2J\phi\sin\tfrac{q_i}{2}i\sigma_2.
\end{align}

\subsection{BdG Hamiltonian}
The BdG Hamiltonian can be diagonalized as
\begin{align}
\hat{H}^{\mathrm{BdG}}=\sum_{\bm{k}}E_{\bm{k}}
\begin{pmatrix}
\hat{\gamma}_{\bm{k}\uparrow}^\dagger & \hat{\gamma}_{-\bm{k}\downarrow}
\end{pmatrix}
\sigma_3
\begin{pmatrix}
\hat{\gamma}_{\bm{k}\uparrow}\\ \hat{\gamma}_{-\bm{k}\downarrow}^\dagger 
\end{pmatrix}+C,
\end{align}
where
\begin{align}
&\begin{pmatrix}
\hat{\gamma}_{\bm{k}\uparrow}\\ \hat{\gamma}_{-\bm{k}\downarrow}^\dagger 
\end{pmatrix}
\coloneqq u_{\bm{k}}^\dagger
\hat{\Psi}_{\bm{k}},\\
&u_{\bm{k}}\coloneqq\frac{1}{\sqrt{|\Delta|^2+(E_{\bm{k}}+\xi_{\bm{k}})^2}}
\begin{pmatrix}
E_{\bm{k}}+\xi_{\bm{k}}&-\Delta\\
\Delta&E_{\bm{k}}+\xi_{\bm{k}}
\end{pmatrix}.
\end{align}
It follows that $\hat{\gamma}_{\bm{k}\sigma}^\dagger(t)=\hat{\gamma}_{\bm{k}\sigma}^\dagger e^{iE_{\bm{k}}t}$ in the Heisenberg picture.

The annihilation operators of electrons can be expressed in terms of operators for Bogoloiubov quasi-particles:
\begin{align}
&\hat{c}_{\bm{k}\uparrow}=\frac{(E_{\bm{k}}+\xi_{\bm{k}})\hat{\gamma}_{\bm{k}\uparrow}-\Delta\hat{\gamma}_{-\bm{k}\downarrow}^\dagger}{\sqrt{|\Delta|^2+(E_{\bm{k}}+\xi_{\bm{k}})^2}},\\
&\hat{c}_{\bm{k}\downarrow}=\frac{(E_{\bm{k}}+\xi_{\bm{k}})\hat{\gamma}_{\bm{k}\downarrow}+\Delta\hat{\gamma}_{-\bm{k}\uparrow}^\dagger}{\sqrt{|\Delta|^2+(E_{\bm{k}}+\xi_{\bm{k}})^2}}.
\end{align}
From these expressions, we find
\begin{align}
\langle\hat{c}_{-\bm{k}\downarrow}\hat{c}_{\bm{k}\uparrow}\rangle=-\frac{\Delta}{2E_{\bm{k}}},\quad
\langle\hat{c}_{\bm{k}\sigma}^\dagger\hat{c}_{\bm{k}\sigma}\rangle=\frac{|\Delta|^2}{|\Delta|^2+(E_{\bm{k}}+\xi_{\bm{k}})^2}=\frac{E_{\bm{k}}-\xi_{\bm{k}}}{2E_{\bm{k}}}.
\end{align}
In the derivation, we used
\begin{align}
\frac{E_{\bm{k}}+\xi_{\bm{k}}}{|\Delta|^2+(E_{\bm{k}}+\xi_{\bm{k}})^2}=\frac{1}{2E_{\bm{k}}},\quad
\frac{|\Delta|^2}{|\Delta|^2+(E_{\bm{k}}+\xi_{\bm{k}})^2}=\frac{E_{\bm{k}}-\xi_{\bm{k}}}{2E_{\bm{k}}}.
\end{align}

\subsection{Green's function}
Let us define the (matrix-valued) Green function
\begin{align}
G(x,x')&\coloneqq-i\langle \mathcal{T}\hat{\Psi}_{\bm{x}}(t)\hat{\Psi}_{\bm{x}'}^\dagger(t')\rangle=\frac{1}{V}\sum_{\bm{q}}\int\frac{d\omega}{2\pi}e^{i\bm{q}\cdot(\bm{x}-\bm{x}')-i\omega(t-t')}G_q
\end{align}
for the Nambu spinor
\begin{align}
\hat{\Psi}_{\bm{x}}\coloneqq\begin{pmatrix}
\hat{c}_{\bm{x}\uparrow}\\ \hat{c}_{-\bm{x}\downarrow}^\dagger 
\end{pmatrix}.
\end{align}
In the Fourier space, it reads as
\begin{align}
G_{\bm{q}}(t)&\coloneqq-i\langle \mathcal{T}\hat{\Psi}_{\bm{q}}(t)\hat{\Psi}_{\bm{q}}^\dagger\rangle\notag\\
&=-i\theta(t)\frac{e^{-iE_{\bm{q}}t}}{2E_{\bm{q}}}
\begin{pmatrix}
E_{\bm{q}}+\xi_{\bm{q}}&\Delta\\
\Delta&E_{\bm{q}}-\xi_{\bm{q}}
\end{pmatrix}
+i\theta(-t)\frac{e^{iE_{\bm{q}}t}}{2E_{\bm{q}}}
\begin{pmatrix}
E_{\bm{q}}-\xi_{\bm{q}}&-\Delta\\
-\Delta&E_{\bm{q}}+\xi_{\bm{q}}
\end{pmatrix}\notag\\
&=-i\theta(t)e^{-iE_{\bm{q}}t}\frac{E_{\bm{q}}\sigma_0+\xi_{\bm{q}}\sigma_3+\Delta\sigma_1}{2E_{\bm{q}}}
+i\theta(-t)e^{iE_{\bm{q}}t}\frac{E_{\bm{q}}\sigma_0-\xi_{\bm{q}}\sigma_3-\Delta\sigma_1}{2E_{\bm{q}}}
\end{align}
and
\begin{align}
G_q&\coloneqq \int dte^{i\omega t}G_{\bm{q}}(t)\notag\\
&=\frac{1}{\omega-E_{\bm{q}}+i\delta}
\frac{E_{\bm{q}}\sigma_0+\xi_{\bm{q}}\sigma_3+\Delta\sigma_1}{2E_{\bm{q}}}
+\frac{1}{\omega +E_{\bm{q}}-i\delta}
\frac{E_{\bm{q}}\sigma_0-\xi_{\bm{q}}\sigma_3-\Delta\sigma_1}{2E_{\bm{q}}}\notag\\
&=\frac{\omega\sigma_0+\xi_{\bm{q}}\sigma_3+\Delta\sigma_1}{\omega^2-E_{\bm{q}}^2+i\delta}.
\end{align}
We have
\begin{align}
G_{\bm{q}}(\pm\eta)=\int\frac{d\omega}{2\pi}G_{q}e^{\mp i\omega\eta }=-i\frac{\pm E_{\bm{q}}\sigma_0+\xi_{\bm{q}}\sigma_3+\Delta\sigma_1}{2E_{\bm{q}}}.
\end{align}
\begin{align}
&G_{0q}^{-1}=\omega\sigma_0-\xi_{\bm{q}}\sigma_3,\\
&G_q^{-1}=\omega\sigma_0-\xi_{\bm{q}}\sigma_3-\Delta\sigma_1,\\
&\Sigma(q)=G_{0q}^{-1}-G_q^{-1}=\Delta\sigma_1.
\end{align}

\subsection{Generalized Ward identity}
We define the vertex functions $\Gamma_0$ and $\Gamma_i$ by
\begin{align}
&\Lambda_0(x,y,\bm{z},t_z)\coloneqq\langle \mathcal{T}\hat{n}_{\bm{z}}(t_z)\hat{\Psi}_{\bm{x}}(t_x)\hat{\Psi}_{\bm{y}}^\dagger(t_y)\rangle=-\sum_{\bm{x}',\bm{y}'}\int dt_x'dt_y'G(x,x')\Gamma_0(x',y',\bm{z},t_z)G(y',y),\\
&\Lambda_i(x,y,\bm{z}+\tfrac{1}{2}\bm{e}_i,t_z)\coloneqq\langle \mathcal{T}\hat{j}_{\bm{z},\bm{z}+\bm{e}_i}(t_z)\hat{\Psi}_{\bm{x}}(t_x)\hat{\Psi}_{\bm{y}}^\dagger(t_y)\rangle=-\sum_{\bm{x}',\bm{y}'}\int dt_x'dt_y'G(x,x')\Gamma_i(x',y',\bm{z}+\tfrac{1}{2}\bm{e}_i,t_z)G(y',y).
\end{align}
Using the continuity equation, we find
\begin{align}
&\partial_{t_z}\Lambda_0(x,y,\bm{z},t_z)+\sum_{i=1}^d\big[\Lambda_i(x,y,\bm{z}+\tfrac{1}{2}\bm{e}_i,t_z)-\Lambda_i(x,y,\bm{z}-\tfrac{1}{2}\bm{e}_i,t_z)\big]\notag\\
&=\delta(t_z-t_y)\delta_{\bm{z},\bm{y}}G(x,z)i\sigma_3\notag-\delta(t_z-t_x)\delta_{\bm{z},\bm{x}}i\sigma_3G(z,y)\\
&=-\sum_{\bm{x}',\bm{y}'}\int dt_xdt_yG(x,x')\big[\partial_{t_z}\Gamma_0(x',y',\bm{z},t_z)+\Gamma_i(x',y',\bm{z}+\tfrac{1}{2}\bm{e}_i,t_z)-\Gamma_i(x',y',\bm{z}-\tfrac{1}{2}\bm{e}_i,t_z)\big]G(y',y).
\end{align}
In the derivation, we used
\begin{align}
&\delta(t_z-t_x)\langle\mathcal{T}\big[\hat{n}_{\bm{z}}(t_x),\hat{\Psi}_{\bm{x}}(t_x)\big]\hat{\Psi}_{\bm{y}}^\dagger(t_y)\rangle+\delta(t_z-t_y)\langle\mathcal{T}\hat{\Psi}_{\bm{x}}(t_x)\big[\hat{n}_{\bm{z}}(t_y),\hat{\Psi}_{\bm{y}}^\dagger(t_y)\big]\rangle\notag\\
&=-\delta(t_z-t_x)\delta_{\bm{z},\bm{x}}\sigma_3\langle\mathcal{T}\hat{\Psi}_{\bm{x}}(t_x)\hat{\Psi}_{\bm{y}}^\dagger(t_y)\rangle+\delta(t_z-t_y)\delta_{\bm{z},\bm{y}}\langle\mathcal{T}\hat{\Psi}_{\bm{x}}(t_x)\hat{\Psi}_{\bm{y}}^\dagger(t_y)\rangle\sigma_3\notag\\
&=-\delta(t_z-t_x)\delta_{\bm{z},\bm{x}}i\sigma_3G(x,y)+\delta(t_z-t_y)\delta_{\bm{z},\bm{y}}G(x,y)i\sigma_3.
\end{align}
Introducing the Fourier transformation by 
\begin{align}
&\Gamma_0(x',y',\bm{z},t_z)=\int \frac{d\omega'}{2\pi}\int \frac{d\omega}{2\pi}\frac{1}{V^2}\sum_{\bm{p},\bm{q}}\Gamma_{0,p+q,p}e^{i\bm{p}\cdot(\bm{x}'-\bm{y}')+i\bm{q}\cdot(\bm{x}'-\bm{z})-i\omega'(t_x'-t_y')-i\omega(t_x'-t_z)},\\
&\Gamma_i(x',y',\bm{z}+\tfrac{1}{2}\bm{e}_i,t_z)=\int \frac{d\omega'}{2\pi}\int \frac{d\omega}{2\pi}\frac{1}{V^2}\sum_{\bm{p},\bm{q}}\Gamma_{i,p+q,p}e^{i\bm{p}\cdot(\bm{x}'-\bm{y}')+i\bm{q}\cdot(\bm{x}'-\bm{z}-\frac{1}{2}\bm{e}_i)-i\omega'(t_x'-t_y')-i\omega(t_x'-t_z)},
\end{align}
we find
\begin{align}
&\langle \mathcal{T}\hat{n}_{\bm{z}}(t_z)\hat{\Psi}_{\bm{x}}(t_x)\hat{\Psi}_{\bm{y}}^\dagger(t_y)\rangle\notag\\
&=-\int \frac{d\omega'}{2\pi}\int \frac{d\omega}{2\pi}\frac{1}{V^2}\sum_{\bm{p},\bm{q}}
G_{p+q}\Gamma_{0,p+q,p}G_pe^{i(\bm{p}+\bm{q})\cdot\bm{x}-i(\omega'+\omega)(t_x-t_z)}e^{i\bm{p}\cdot(\bm{z}-\bm{y})-i\omega'(t_z-t_y)}
\end{align}
and
\begin{align}
&\langle \mathcal{T}\hat{j}_{\bm{z},\bm{z}+\bm{e}_i}(t_z)\hat{\Psi}_{\bm{x}}(t_x)\hat{\Psi}_{\bm{y}}^\dagger(t_y)\rangle\notag\\
&=-\int \frac{d\omega'}{2\pi}\int \frac{d\omega}{2\pi}\frac{1}{V^2}\sum_{\bm{p},\bm{q}}
G_{p+q}\Gamma_{i,p+q,p}G_pe^{i(\bm{p}+\bm{q})\cdot(\bm{x}-\bm{z}-\frac{1}{2}\bm{e}_i)-i(\omega'+\omega)(t_x-t_z)}e^{i\bm{p}\cdot(\bm{z}+\frac{1}{2}\bm{e}_i-\bm{y})-i\omega'(t_z-t_y)}.
\end{align}
Here, $p=(\omega',\bm{p})$.
Using these expressions, we arrive at the generalized Ward identity:
\begin{align}
\sum_{i=1}^d2\sin\tfrac{q_i}{2}\Gamma_{i,p+q,p}-\omega\Gamma_{0,p+q,p}=\sigma_3G_p^{-1}-G_{p+q}^{-1}\sigma_3.
\end{align}
Writing 
\begin{align}
\gamma_{i,\bm{k}+\bm{q},\bm{k}}^{(0)}\coloneqq v_{i,\bm{k}+\frac{\bm{q}}{2}}\sigma_0,\quad\gamma_{0,\bm{p}+\bm{q},\bm{p}}^{(0)}=\sigma_3,
\end{align}
we find
\begin{align}
&\sum_{i=1}^d2\sin\tfrac{q_i}{2}\gamma_{i,\bm{p}+\bm{q},\bm{p}}^{(0)}-\omega\gamma_{0,\bm{p}+\bm{q},\bm{p}}^{(0)}=\sigma_3G_{0p}^{-1}-G_{0p+q}^{-1}\sigma_3.
\end{align}

\subsection{Current correlation function} 
We define the time-ordered current correlation function by
\begin{align}
P_{\mu\nu}(\bm{q},t)\coloneqq -i\langle \mathcal{T}\hat{J}_{\mu,\bm{q}}(t)\hat{J}_{\nu,-\bm{q}}\rangle.
\end{align}
We find
\begin{align}
P_{ij}(q)&\coloneqq -\frac{i}{V}\int dt e^{i\omega t}\langle \mathcal{T}\hat{J}_{i,\bm{q}}(t)\hat{J}_{j,-\bm{q}}(0)\rangle\notag\\
&=-\frac{i}{V}\int dt e^{i\omega t}\sum_{\bm{p}}\langle \mathcal{T}(\hat{\Psi}_{\bm{p}})_l^\dagger(t+\eta)(\gamma_{i,\bm{p}+\bm{q},\bm{p}})_{lm}(\hat{\Psi}_{\bm{p}+\bm{q}})_m(t)\hat{J}_{j,-\bm{q}}(0)\rangle\notag\\
&=\frac{i}{V}\int dt e^{i\omega t}\sum_{\bm{p}}(\gamma_{i,\bm{p}+\bm{q},\bm{p}})_{lm}\langle \mathcal{T}\hat{J}_{j,-\bm{q}}(0)(\hat{\Psi}_{\bm{p}+\bm{q}})_m(t)(\hat{\Psi}_{\bm{p}})_l^\dagger(t+\eta)\rangle\notag\\
&=\frac{i}{V}\int dt e^{i\omega t}\sum_{\bm{p}}\frac{1}{V}\sum_{\bm{x},\bm{y},\bm{z}}\text{tr}[\gamma_{i,\bm{p}+\bm{q},\bm{p}}\langle \mathcal{T}\hat{j}_{\bm{z},\bm{z}+\bm{e}_j}(0)\hat{\Psi}_{\bm{x}}(t)\hat{\Psi}_{\bm{y}}^\dagger(t+\eta)\rangle]e^{-i(\bm{p}+\bm{q})\cdot\bm{x}}e^{i\bm{p}\cdot\bm{y}}e^{i\bm{q}\cdot(\bm{z}+\frac{1}{2}\bm{e}_j)}\notag\\
&=-\int \frac{d\omega'}{2\pi}\frac{i}{V}\sum_{\bm{p}}\text{tr}[\gamma_{i,\bm{p}+\bm{q},\bm{p}}G_{p+q}\Gamma_{j,p+q,p}G_p]e^{i\omega'\eta},
\end{align}
where we used
\begin{align}
&\langle \mathcal{T}\hat{j}_{\bm{z},\bm{z}+\bm{e}_j}(0)\hat{\Psi}_{\bm{x}}(t)\hat{\Psi}_{\bm{y}}^\dagger(t+\eta)\rangle\notag\\
&=-\int \frac{d\omega'}{2\pi}\int \frac{d\omega}{2\pi}\frac{1}{V^2}\sum_{\bm{p},\bm{q}}
G_{p+q}\Gamma_{j,p+q,p}G_pe^{i(\bm{p}+\bm{q})\cdot(\bm{x}-\bm{z}-\frac{1}{2}\bm{e}_j)-i\omega t}e^{i\bm{p}\cdot(\bm{z}+\frac{1}{2}\bm{e}_j-\bm{y})+i\omega'\eta}.
\end{align}

\begin{align}
\sum_{j=1}^dP_{0j}(q)2\sin\tfrac{q_j}{2} -P_{00}(q)\omega
&=-\int \frac{d\omega'}{2\pi}\frac{i}{V}\sum_{\bm{p}}\text{tr}[\sigma_3(G_{p+q}\sigma_3-\sigma_3G_p)]e^{i\omega'\eta}\notag\\
&=-\int \frac{d\omega'}{2\pi}\frac{i}{V}\sum_{\bm{p}}\text{tr}[G_{p+q}-G_p]e^{i\omega'\eta}=0.
\end{align}

\begin{align}
\sum_{j=1}^dP_{ij}(q)2\sin\tfrac{q_j}{2}-P_{i0}(q)\omega
&=-\int \frac{d\omega'}{2\pi}\frac{i}{V}\sum_{\bm{p}}\text{tr}[\gamma_{i,\bm{p}+\bm{q},\bm{p}}(G_{p+q}\sigma_3-\sigma_3G_p)]e^{i\omega'\eta}\notag\\
&=-\int \frac{d\omega'}{2\pi}\frac{i}{V}\sum_{\bm{p}}\text{tr}[(\sigma_3\gamma_{i,\bm{p},\bm{p}-\bm{q}}e^{-i\omega\eta}-\gamma_{i,\bm{p}+\bm{q},\bm{p}}\sigma_3)G_p]e^{i\omega'\eta}\notag\\
&=-\frac{1}{V}\sum_{\bm{p}}\text{tr}\left[\frac{\big((v_{i,\bm{p}-\frac{\bm{q}}{2}}-v_{i,\bm{p}+\frac{\bm{q}}{2}})\sigma_3-4J\phi\sin\tfrac{q_i}{2}\sigma_1\big)\big(- E_{\bm{q}}\sigma_0+\xi_{\bm{q}}\sigma_3-\phi U_{\text{tot}}\sigma_1\big)}{2E_{\bm{q}}}\right]\notag\\
&=-2\sin\tfrac{q_i}{2}\frac{1}{V}\sum_{\bm{p}}\frac{-4t\cos p_i\xi_{\bm{q}}+4U_{\text{tot}}J|\phi|^2}{2E_{\bm{q}}}\notag\\
&=-2\sin\tfrac{q_i}{2}\left(\frac{1}{V}\sum_{\bm{p}}2t\cos p_i\langle \hat{n}_{\bm{p}}\rangle+4J|\phi|^2\right).
\end{align}

\subsection{Bethe-Salpeter equation}
The Bethe-Salpeter equation
\begin{align}
\Gamma_{\mu,p+q,p}=\gamma_{\mu,\bm{p}+\bm{q},\bm{p}}^{(0)}-U_{\text{tot}}\int\frac{dk_0}{2\pi i}\frac{1}{V}\sum_{\bm{k}}\sigma_3G_{k+q}\Gamma_{\mu,k+q,k}G_k\sigma_3
\end{align}
is consistent with Ward identities
\begin{align}
&\sum_{i=1}^d2\sin\tfrac{q_i}{2}\Gamma_{i,p+q,p}-\omega\Gamma_{0,p+q,p}=\sigma_3G_p^{-1}-G_{p+q}^{-1}\sigma_3,\\
&\sum_{i=1}^d2\sin\tfrac{q_i}{2}\gamma_{i,\bm{p}+\bm{q},\bm{p}}^{(0)}-\omega\gamma_{0,\bm{p}+\bm{q},\bm{p}}^{(0)}=\sigma_3G_{0p}^{-1}-G_{0p+q}^{-1}\sigma_3.
\end{align}
Indeed, we have
\begin{align}
&\sigma_3G_p^{-1}-G_{p+q}^{-1}\sigma_3\notag\\
&=\sum_{i=1}^d2\sin\tfrac{q_i}{2}\gamma_{i,\bm{p}+\bm{q},\bm{p}}^{(0)}-\omega\gamma_{0,\bm{p}+\bm{q},\bm{p}}^{(0)}-U_{\text{tot}}\int\frac{dk_0}{2\pi i}\frac{1}{V}\sum_{\bm{k}}\sigma_3G_{k+q}[\sum_{i=1}^d2\sin\tfrac{q_i}{2}\Gamma_{i,k+q,k}-\omega\Gamma_{0,k+q,k}]G_k\sigma_3\notag\\
&=\sigma_3G_{0p}^{-1}-G_{0p+q}^{-1}\sigma_3-U_{\text{tot}}\int\frac{dk_0}{2\pi i}\frac{1}{V}\sum_{\bm{k}}\sigma_3G_{k+q}[\sigma_3G_k^{-1}-G_{k+q}^{-1}\sigma_3]G_k\sigma_3\notag\\
&=\sigma_3G_{0p}^{-1}-G_{0p+q}^{-1}\sigma_3-U_{\text{tot}}\int\frac{dk_0}{2\pi i}\frac{1}{V}\sum_{\bm{k}}[\sigma_3G_{k+q}-G_k\sigma_3]\notag\\
&=\sigma_3G_{0p}^{-1}-G_{0p+q}^{-1}\sigma_3-U_{\text{tot}}\frac{1}{V}\sum_{\bm{k}}\left[\frac{\pm E_{\bm{k}}\sigma_3+\xi_{\bm{k}}\sigma_0+\Delta i\sigma_2}{2E_{\bm{k}}}-\frac{\pm E_{\bm{k}}\sigma_3+\xi_{\bm{k}}\sigma_0-\Delta i\sigma_2}{2E_{\bm{k}}}\right]\notag\\
&=\sigma_3G_{0p}^{-1}-G_{0p+q}^{-1}\sigma_3-2i\sigma_2U_{\text{tot}}\frac{1}{V}\sum_{\bm{k}}\frac{\Delta}{2E_{\bm{k}}}\notag\\
&=\sigma_3G_{0p}^{-1}-G_{0p+q}^{-1}\sigma_3-2i\sigma_2\Delta.
\end{align}
In the derivation, we used
\begin{align}
\int\frac{d\omega}{2\pi i}G_{q}e^{\mp i\omega\eta }=\frac{\pm E_{\bm{q}}\sigma_0+\xi_{\bm{q}}\sigma_3+\Delta\sigma_1}{2E_{\bm{q}}}.
\end{align}

\clearpage

\section{Spinless model}
\label{app:spinless}

Here let us summarize the symmetry and conserved current in the spinless model. The Hamiltonian is given by
\begin{align}
\hat{H}&\coloneqq -\sum_{\bm{x}}\Big(\sum_{i=1,2}t(e^{-iA_i}\hat{c}_{\bm{x}+\bm{e}_i}^\dagger \hat{c}_{\bm{x}}+e^{iA_i}\hat{c}_{\bm{x}}^\dagger \hat{c}_{\bm{x}+\bm{e}_i})+\mu\hat{n}_{\bm{x}}\Big)-\frac{U_0}{2}\sum_{\bm{x}}\sum_{i=1,2}\hat{n}_{\bm{x}}\hat{n}_{\bm{x}+\bm{e}_i}\notag\\
&\quad-\frac{J}{8}\sum_{\bm{x}}
i\hat{n}_{\bm{x}}\Big[(e^{i(\theta_2-\theta_1)}\hat{c}_{\bm{x}+\bm{e}_1}^\dagger \hat{c}_{\bm{x}+\bm{e}_2}-e^{-i(\theta_2-\theta_1)}\hat{c}_{\bm{x}+\bm{e}_2}^\dagger \hat{c}_{\bm{x}+\bm{e}_1})+(e^{-i(\theta_1+\theta_2)}\hat{c}_{\bm{x}+\bm{e}_2}^\dagger \hat{c}_{\bm{x}-\bm{e}_1}-e^{i(\theta_1+\theta_2)}\hat{c}_{\bm{x}-\bm{e}_1}^\dagger \hat{c}_{\bm{x}+\bm{e}_2})\notag\\
&\quad\quad\quad\quad+(e^{i(\theta_1-\theta_2)}\hat{c}_{\bm{x}-\bm{e}_1}^\dagger \hat{c}_{\bm{x}-\bm{e}_2}-e^{-i(\theta_1-\theta_2)}\hat{c}_{\bm{x}-\bm{e}_2}^\dagger \hat{c}_{\bm{x}-\bm{e}_1})+(e^{i(\theta_1+\theta_2)}\hat{c}_{\bm{x}-\bm{e}_2}^\dagger \hat{c}_{\bm{x}+\bm{e}_1}-e^{-i(\theta_1+\theta_2)}\hat{c}_{\bm{x}+\bm{e}_1}^\dagger \hat{c}_{\bm{x}-\bm{e}_2})\Big].
\end{align}
The local current operators for the U(1) charge $\hat{N}\coloneqq\sum_{\bm{x}}\hat{n}_{\bm{x}}$ are given by
\begin{align}
&\hat{j}_{\bm{x},\bm{x}+\bm{e}_i}\coloneqq it\big(e^{-iA_i}\hat{c}_{\bm{x}+\bm{e}_i}^\dagger \hat{c}_{\bm{x}}-e^{iA_i}\hat{c}_{\bm{x}}^\dagger \hat{c}_{\bm{x}+\bm{e}_i}\big),\\
&\hat{j}_{\bm{x},\bm{x}+\bm{e}_1+\bm{e}_2}\coloneqq
\frac{J}{8}(\hat{n}_{\bm{x}+\bm{e}_2}-\hat{n}_{\bm{x}+\bm{e}_1})(e^{-i(\theta_1+\theta_2)}\hat{c}_{\bm{x}+\bm{e}_1+\bm{e}_2}^\dagger \hat{c}_{\bm{x}}+e^{i(\theta_1+\theta_2)}\hat{c}_{\bm{x}}^\dagger \hat{c}_{\bm{x}+\bm{e}_1+\bm{e}_2}),\\
&\hat{j}_{\bm{x},\bm{x}+\bm{e}_1-\bm{e}_2}\coloneqq\frac{J}{8}(\hat{n}_{\bm{x}+\bm{e}_1}-\hat{n}_{\bm{x}-\bm{e}_2})(e^{i(\theta_2-\theta_1)}\hat{c}_{\bm{x}+\bm{e}_1-\bm{e}_2}^\dagger \hat{c}_{\bm{x}}+e^{-i(\theta_2-\theta_1)}\hat{c}_{\bm{x}}^\dagger \hat{c}_{\bm{x}+\bm{e}_1-\bm{e}_2}).
\end{align}
The continuity equation is
\begin{align}
\partial_t\hat{n}_{\bm{x}}&=-i[\hat{n}_{\bm{x}},\hat{H}]\notag\\
&=-\sum_{i=1,2}\big(\hat{j}_{\bm{x},\bm{x}+\bm{e}_i}-\hat{j}_{\bm{x}-\bm{e}_i,\bm{x}}\big)-(\hat{j}_{\bm{x},\bm{x}+\bm{e}_1+\bm{e}_2}-\hat{j}_{\bm{x-\bm{e}_1-\bm{e}_2},\bm{x}})-(\hat{j}_{\bm{x},\bm{x}+\bm{e}_1-\bm{e}_2}-\hat{j}_{\bm{x-\bm{e}_1+\bm{e}_2},\bm{x}}).\label{ceq2}
\end{align}
We introduce the Fourier transformation by
\begin{align}
&\hat{N}_{\bm{q}}\coloneqq \sum_{\bm{x}}e^{-i\bm{q}\cdot\bm{x}}\hat{n}_{\bm{x}},\\
&\hat{J}_{\bm{q},\bm{e}_i}\coloneqq \sum_{\bm{x}}e^{-i\bm{q}\cdot(\bm{x}+\frac{1}{2}\bm{e}_i)}\hat{j}_{\bm{x},\bm{x}+\bm{e}_i},\\
&\hat{J}_{\bm{q},\bm{e}_1+\bm{e}_2}\coloneqq \sum_{\bm{x}}e^{-i\bm{q}\cdot(\bm{x}+\frac{1}{2}(\bm{e}_1+\bm{e}_2))}\hat{j}_{\bm{x},\bm{x}+\bm{e}_1+\bm{e}_2},\\
&\hat{J}_{\bm{q},\bm{e}_1-\bm{e}_2}\coloneqq \sum_{\bm{x}}e^{-i\bm{q}\cdot(\bm{x}+\frac{1}{2}(\bm{e}_1-\bm{e}_2))}\hat{j}_{\bm{x},\bm{x}+\bm{e}_1-\bm{e}_2}.
\end{align}
Then the continuity equation becomes
\begin{align}
[\hat{N}_{\bm{q}},\hat{H}]
&=\sum_{i=1,2}2\sin\tfrac{q_i}{2}\,\hat{J}_{\bm{q},\bm{e}_i}+2\sin\tfrac{q_1+q_2}{2}\,\hat{J}_{\bm{q},\bm{e}_1+\bm{e}_2}+2\sin\tfrac{q_1-q_2}{2}\,\hat{J}_{\bm{q},\bm{e}_1-\bm{e}_2}.
\label{ceq2}
\end{align}
In the $|\bm{q}|\rightarrow0$ limit, 
\begin{align}
[\hat{N}_{\bm{q}},\hat{H}]=\sum_{i=1,2}q_i\hat{J}_{i,\bm{q}},
\label{ceq3}
\end{align}
where
\begin{align}
&\hat{J}_{1,\bm{q}}\coloneqq\hat{J}_{\bm{q},\bm{e}_1}+\hat{J}_{\bm{q},\bm{e}_1+\bm{e}_2}+\hat{J}_{\bm{q},\bm{e}_1-\bm{e}_2},\\
&\hat{J}_{2,\bm{q}}\coloneqq\hat{J}_{\bm{q},\bm{e}_1}+\hat{J}_{\bm{q},\bm{e}_1+\bm{e}_2}-\hat{J}_{\bm{q},\bm{e}_1-\bm{e}_2}.
\end{align}

\clearpage

\section{Correlation functions}
\label{app:cf}

The time-ordered current correlation is defined as
\begin{align}
\mathcal{P}_{\mu\nu}(q)&\coloneqq-\frac{i}{V}\int dte^{i\omega t}\langle\mathcal{T}\hat{J}_{\mu,\bm{q}}(t)\hat{J}_{\nu,-\bm{q}}(0)\rangle\notag\\
&=-\frac{i}{V}\int dt\Big(\theta(t)\langle\hat{J}_{\mu,\bm{q}}(t)\hat{J}_{\nu,-\bm{q}}(0)\rangle+\theta(-t)\langle\hat{J}_{\nu,-\bm{q}}(0)\hat{J}_{\mu,\bm{q}}(t)\rangle\Big)\notag\\
&=-\frac{i}{V}\sum_n\int dt\Big(\theta(t)\langle0|\hat{J}_{\mu,\bm{q}}|n\rangle\langle n|\hat{J}_{\nu,-\bm{q}}|0\rangle e^{i(\omega-E_n+E_0)t}+\theta(-t)\langle0|\hat{J}_{\nu,-\bm{q}}|n\rangle\langle n|\hat{J}_{\mu,\bm{q}}|0\rangle e^{i(\omega+E_n-E_0)t}\Big)\notag\\
&=\frac{1}{V}\sum_n\Big(\frac{\langle0|\hat{J}_{\mu,\bm{q}}|n\rangle\langle n|\hat{J}_{\nu,-\bm{q}}|0\rangle }{\omega-E_n+E_0+i\delta}-\frac{\langle0|\hat{J}_{\nu,-\bm{q}}|n\rangle\langle n|\hat{J}_{\mu,\bm{q}}|0\rangle}{\omega+E_n-E_0-i\delta}\Big)\notag\\
&=\frac{1}{V}\sum_nP\Big(\frac{\langle0|\hat{J}_{\mu,\bm{q}}|n\rangle\langle n|\hat{J}_{\nu,-\bm{q}}|0\rangle}{\omega-E_n+E_0}-\frac{\langle0|\hat{J}_{\nu,-\bm{q}}|n\rangle\langle n|\hat{J}_{\mu,\bm{q}}|0\rangle}{\omega+E_n-E_0}\Big)\notag\\
&\quad-\frac{i\pi}{V}\sum_n\Big(\langle0|\hat{J}_{\mu,\bm{q}}|n\rangle\langle n|\hat{J}_{\nu,-\bm{q}}|0\rangle\delta(\omega-E_n+E_0)+\langle0|\hat{J}_{\nu,-\bm{q}}|n\rangle\langle n|\hat{J}_{\mu,\bm{q}}|0\rangle\delta(\omega+E_n-E_0)\Big)
\end{align}

The retarded current correlation is defined as
\begin{align}
\mathcal{R}_{\mu\nu}(q)&\coloneqq-\frac{i}{V}\int dte^{i\omega t}\theta(t)\langle[\hat{J}_{\mu,\bm{q}}(t),\hat{J}_{\nu,-\bm{q}}(0)]\rangle\notag\\
&=-\frac{i}{V}\int dt\theta(t)\Big(\langle\hat{J}_{\mu,\bm{q}}(t)\hat{J}_{\nu,-\bm{q}}(0)\rangle-\langle\hat{J}_{\nu,-\bm{q}}(0)\hat{J}_{\mu,\bm{q}}(t)\rangle\Big)\notag\\
&=-\frac{i}{V}\sum_n\int dt\theta(t)\Big(\langle0|\hat{J}_{\mu,\bm{q}}|n\rangle\langle n|\hat{J}_{\nu,-\bm{q}}|0\rangle e^{i(\omega-E_n+E_0)t}-\langle0|\hat{J}_{\nu,-\bm{q}}|n\rangle\langle n|\hat{J}_{\mu,\bm{q}}|0\rangle e^{i(\omega+E_n-E_0)t}\Big)\notag\\
&=\frac{1}{V}\sum_n\Big(\frac{\langle0|\hat{J}_{\mu,\bm{q}}|n\rangle\langle n|\hat{J}_{\nu,-\bm{q}}|0\rangle }{\omega-E_n+E_0+i\delta}-\frac{\langle0|\hat{J}_{\nu,-\bm{q}}|n\rangle\langle n|\hat{J}_{\mu,\bm{q}}|0\rangle}{\omega+E_n-E_0+i\delta}\Big)\notag\\
&=\frac{1}{V}\sum_nP\Big(\frac{\langle0|\hat{J}_{\mu,\bm{q}}|n\rangle\langle n|\hat{J}_{\nu,-\bm{q}}|0\rangle}{\omega-E_n+E_0}-\frac{\langle0|\hat{J}_{\nu,-\bm{q}}|n\rangle\langle n|\hat{J}_{\mu,\bm{q}}|0\rangle}{\omega+E_n-E_0}\Big)\notag\\
&\quad-\frac{i\pi}{V}\sum_n\Big(\langle0|\hat{J}_{\mu,\bm{q}}|n\rangle\langle n|\hat{J}_{\nu,-\bm{q}}|0\rangle\delta(\omega-E_n+E_0)-\langle0|\hat{J}_{\nu,-\bm{q}}|n\rangle\langle n|\hat{J}_{\mu,\bm{q}}|0\rangle\delta(\omega+E_n-E_0)\Big)
\end{align}

Therefore,
\begin{align}
&\text{Re}\mathcal{R}_{\mu\nu}(q)=\text{Re}\mathcal{P}_{\mu\nu}(q)=\frac{1}{V}\sum_nP\Big(\frac{\langle0|\hat{J}_{\mu,\bm{q}}|n\rangle\langle n|\hat{J}_{\nu,-\bm{q}}|0\rangle}{\omega-E_n+E_0}-\frac{\langle0|\hat{J}_{\nu,-\bm{q}}|n\rangle\langle n|\hat{J}_{\mu,\bm{q}}|0\rangle}{\omega+E_n-E_0}\Big),\\
&\text{Im}\mathcal{R}_{\mu\nu}(q)=\text{sign}(\omega)\text{Im}\mathcal{P}_{\mu\nu}(q)=-\frac{\pi}{V}\Big(\langle0|\hat{J}_{\mu,\bm{q}}|n\rangle\langle n|\hat{J}_{\nu,-\bm{q}}|0\rangle\delta(\omega-E_n+E_0)-\langle0|\hat{J}_{\nu,-\bm{q}}|n\rangle\langle n|\hat{J}_{\mu,\bm{q}}|0\rangle\delta(\omega+E_n-E_0)\Big).
\end{align}

\end{document}